\documentclass{emulateapj}
\pdfoutput=1
\usepackage{natbib}
\usepackage{times}
\usepackage{graphicx}
\usepackage{amsmath}

\usepackage[colorlinks=true,citecolor=black,filecolor=blue,linkcolor=blue,urlcolor=blue,pdftex]{hyperref}

\begin{document}

\shorttitle{Gamma Rays from the Milky Way, Part I}
\shortauthors{Kistler}

\title{Gamma Rays, Electrons, Hard X-Rays, and the Central Parsec of the Milky Way}

\author{Matthew D. Kistler}
%\email{kistler@stanford.edu}
\affiliation{Kavli Institute for Particle Astrophysics and Cosmology, Stanford University, SLAC National Accelerator Laboratory, Menlo Park, CA 94025}

%\date{November 25, 2016}

\begin{abstract}
The complex interplay of processes at the Galactic Center is at the heart of numerous past, present, and (likely) future mysteries.  We aim at a more complete understanding of how spectra extending to $>\,$10~TeV result.  We first construct a simplified model to account for the peculiar energy and angular dependence of the intense central parsec photon field.  This allows for calculating anisotropic inverse Compton scattering and mapping gamma-ray extinction due to $\gamma \gamma \!\rightarrow\! e^+ e^-$ attenuation.  Coupling these with a method for evolving electron spectra, we examine several clear and present excesses, including the diffuse hard X-rays seen by {\it NuSTAR} and GeV gamma rays by {\it Fermi}.  We address further applications to cosmic rays, dark matter, neutrinos, and gamma rays from the Center and beyond.

\end{abstract}

%\keywords{gamma-ray burst: general --- galaxies: evolution --- stars: formation }

%--------------------------------------------------------------------%
\section{Introduction}
The Galactic Center (GC) is an arena for astrophysical phenomena unlike any other in our galaxy.  The inner parsec alone is packed with gas streams, dark matter, a puzzling young massive stellar population, and remnants of a long history of star formation, all encircled by a dusty circumnuclear disk \citep{Genzel2010}.  In the middle of this is a supermassive black hole (hereafter Sgr~A$^*$) that typically emits well below Eddington, though with flares that occasionally reach at least into hard X-rays \citep{Barriere2014}.

It is easy to imagine substantial concentrations of photon emission and energetic particles within this region.  Much of the bolometric luminosity has now been identified as originating from the aforementioned massive stars that in turn power the infrared output from dust in the circumnuclear disk (e.g., \citealt{Davidson1992,Krabbe1995,Genzel2010}).  The many matters of central import left to be resolved span the spectrum of photons, cosmic rays, and neutrinos.

Our purpose is to address aspects related to those mysteries that plausibly involve very energetic particles.  These include the origin of the gamut of gamma rays reaching to $>\,$10~TeV \citep{Nolan2012,Aharonian2004,Aharonian2009,Albert2006,Archer2014,Archer2016,Ahnen2016} and bright X-ray emission with non-thermal characteristics.  Such photons, if only for their prime location at the center of the Galactic halo, are of great interest, such as the significant excess of gamma rays at $\sim\! 10$~GeV that has produced considerable excitement (e.g., \citealt{Abazajian2014,Daylan2014,Calore2014,Ajello2016}).

Our focus here is not on providing yet another explanation for such anomalies (not entirely anyway), but rather to better understand the behavior of high-energy particles starting at the Center --- electrons and gamma rays in particular --- via an improved description of the relevant conditions in this unique environment.  For example, recent high spatial resolution infrared data has revealed structures within the central parsec.  These imply a photon background much denser than typically encountered in the Galaxy with variations in the amplitude of each component throughout this region.

A population of electrons, even if their velocity distribution is isotropic, will thus encounter anisotropic photon backgrounds.  Since head-to-head scatterings result in more energy transfer, the resulting inverse Compton spectrum thus depends on the direction to the observer.  Moreover, gamma rays produced via this or other processes can in turn be attenuated by interacting with a background photon to produce an electron-positron pair, the probability of which is dependent on the path taken to the telescope.

We construct a phenomenological energy and angle dependent photon field in the central parsec based on recent infrared data to achieve a basic agreement with the measured broadband spectrum and morphology of the various emissions. This is used to better describe the inverse Compton scattering and $\gamma \gamma \!\rightarrow\! e^+ e^-$ extinction, which have a similar dependence on the geometry of the photon background.

We couple these with a convenient method for calculating time-evolved electron spectra in examining several topics of recent interest.  These include the diffuse hard X-ray emission extending to $>\,$40~keV discovered by {\it NuSTAR} throughout this region that cannot be simply extrapolated from sources prevalent at lower energies \citep{Perez2015,Mori2015}.  We discuss possible attributions, including synchrotron radiation from $\gtrsim\,$100~TeV electrons, and connections to gamma rays.

We also consider contributions from pulsar electrons to the GeV signal seen from the Galactic Center by {\it Fermi} \citep{Acero2015}.  \citet{Kistler2015} extends these techniques in detailing potential TeV gamma-ray signatures of the pulsar wind nebula (PWN) G359.95--0.04 situated at a projected distance of 0.3~pc from Sgr~A$^*$ \citep{Wang2006,Muno2008}.

%%%%%%%%%%%%%%%%%%%%%%%%%%%%%%%%%%%
\begin{figure}[t!]\vspace*{-0.15cm}
\hspace*{-0.1cm}\vspace*{-0.15cm}
\includegraphics[width=1.02\columnwidth,clip=true]{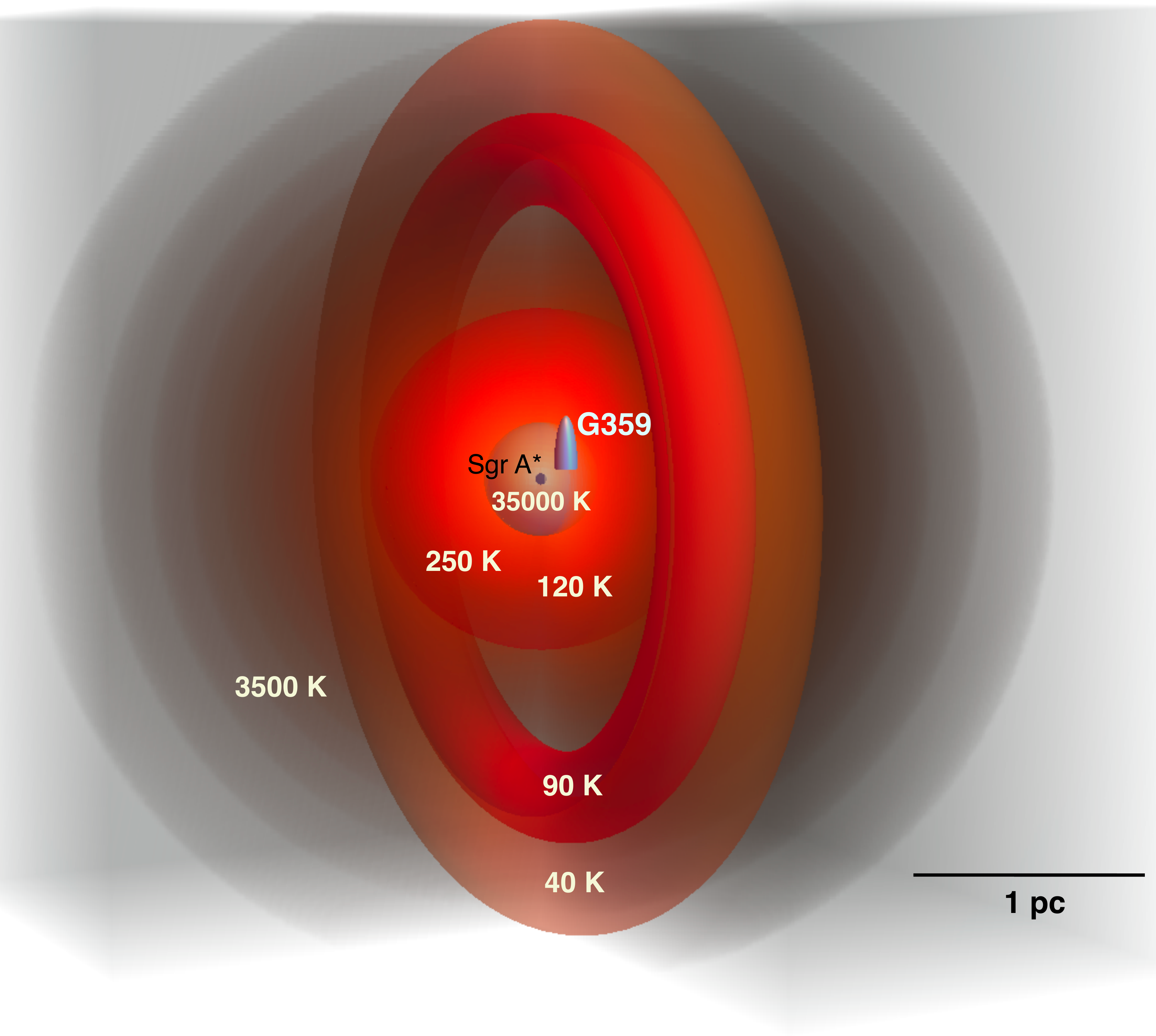}
\caption{An illustration of the geometry of photon emission components from the inner parsec of the Milky Way used in constructing the photon field used throughout this paper, labelled by temperature corresponding to Table~\ref{tab:params}.  The line-of-sight position of PWN G359 is referenced with the blue cone.\\
\label{GCpc}}
\end{figure}
%%%%%%%%%%%%%%%%%%%%%%%%%%%%%%%%%%%

%--------------------------------------------------------------------%
\section{A Portrait of Galactic Center Backgrounds}
The cluster of massive stars at the Galactic Center provides $\gtrsim\! 10^7 \, L_\odot$ of UV photons that drive emission over a broad range of wavelengths.  While UV radiation can be effectively upscattered by GeV electrons, for TeV electrons scattering is suppressed due to the energy dependence of the Klein-Nishina cross section so that infrared emission is their most relevant inverse Compton (IC) target.  Since this cross section depends on the angle between electron and photon, with head-on scattering resulting in a photon with higher energy \citep{Jones1968}, it is of interest to understand the directional variation of the photon field beyond the integrated intensity.

Constructing a first principles model of the energy/angle dependent photon field in the GC would itself be a tremendous achievement.  We rather content ourselves with a satisfactory phenomenological background based on the most recent data.  For easy reference, the component parameters are summarized in Table~\ref{tab:params} and the layout illustrated in Fig.~\ref{GCpc}.

{\it Herschel} has now resolved cold dust in the circumnuclear disk (CND) in the FIR from $70\!-\!500\,\mu$m \citep{Etxaluze2011,Goicoechea2013}.  \citet{Etxaluze2011} also utilized {\it ISO}-LWS data from $46\!-\!180\,\mu$m, which has less angular resolution, to fill in flux from warmer dust.  {\it SOFIA}, with shorter wavelength coverage ($19.7\!-\!37.1\,\mu$m) and sharper resolution, was used to resolve warmer locations of the inner CND in greater detail by \citet{Lau2013}.

We describe these data using two separate rings: one with $T\!=\!90\,$K, $L_{90} \!=\! 2 \times 10^6~L_\odot$, a major radius of $R_{90} \!=\! 1.4\,$pc, and minor radius of $r_{90} \!=\! 0.2\,$pc; the other with $T \!=\! 40\,$K, $L_{40} \!=\! 2 \!\times\! 10^5~L_\odot$, $R_{40} \!=\! 1.7\,$pc, and $r_{40} \!=\! 0.3\,$pc.  The inclination follows the orientation derived in \citet{Lau2013}.  We assume optically thin emission that is uniform throughout the volume with a blackbody spectrum
\begin{equation}
      \frac{dN_i}{d\epsilon_\gamma}  =    \frac{1}{\pi^2 (\hbar c)^3}  \frac{\epsilon_\gamma^2}{e^{\epsilon_\gamma/k_B T_i}-1} 
    \,.
\label{BB}
\end{equation}
This is not formally correct, since optically thin dust has a modified blackbody form with an emissivity $\propto\!\nu^\beta$ and $\beta \!\lesssim\! 2$ that results in a steeper long wavelength tail (e.g., \citealt{Draine2003}).  However, we compensate for this by choosing values for $T$ and $L$ to match the spectral peak for dust of a given temperature (and typically another component becomes more important in the tails).

{\it SOFIA} images display warmer emission nearer the GC \citep{Lau2013}, mostly coinciding with the ionized gas streamers seen in radio (e.g., \citealt{Zhao2010}).  In principle, one can begin from the \citet{Zhao2009} model of Keplerian gas stream orbits to construct a more elaborate model accounting for a heating from a central cluster.  We here assume emission with $T\!=\!120\,$K and $L_{120} \!=\! 1.5\!\times\!10^6~L_\odot$ and approximate the multiple streams with a uniform sphere of radius $R_{120} \!=\! 0.75\,$pc.  The extinction corrected {\it ISO}-SWS spectrum from \citet{Fritz2011}, extending from $2.6\!-\!26\,\mu$m and covering an extended inner portion of the central parsec, as well as radio line measurements (e.g., \citealt{RequenaTorres2012,Mills2013,Smith2014}) also suggest a warmer component that we ascribe to the same volume with $T\!=\!250\,$K and $L_{250} \!=\! 2\!\times\!10^6~L_\odot$.

%
%%%%%%%%%%%%%%%%%%%%%%%%%%%%%
\begin{deluxetable}{rrcc}[t!]
\tabletypesize{\scriptsize}
\tablecaption{\label{tab:params}}
\tablewidth{\columnwidth}
\tablehead{\colhead{$T [K]$} & \colhead{$L~[L_\odot]$} & \colhead{$R$ [pc]} & \colhead{$r$ [pc]}  }
\startdata
\hline \vspace{-0.2cm}\\
%
% \hline
%
35000	& $20\!\times\!10^6$		& 0.25		& --- \\
3500		& $30\!\times\!10^6$		& ---			& --- \\
250		& $2\!\times\!10^6$		& 0.75		& --- \\
120		& $1.5\!\times\!10^6$	& 0.75		& --- \\
90		& $2\!\times\!10^6$		& 1.4			& 0.2 \\
40		& $0.2\!\times\!10^6$	& 1.7			& 0.3 \\
2.73		& CMB				& ---			& --- \\
\hline\vspace{-0.3cm}
\enddata
\tablecomments{Properties of the GC radiation components used here. $R$ refers to the radius of a sphere or major radius of a ring, $r$ to a ring minor radius.\\}
\end{deluxetable}
%%%%%%%%%%%%%%%%%%%%%%%%%%%%%
%

The IR data are consistent with reprocessing of a fraction of the incident UV flux from a $T \!\approx\! 35000\,$K,  $L_{35000} \!\approx\! 2\!\times\!10^7\,L_\odot$ cluster of massive stars at the GC.  \citet{Stostad2015} and \citet{FeldmeierKrause2015} infer a cutoff in the surface brightness by $\sim\!0.5\,$pc for this population, which we approximate with a sphere of $R_{35000} \!=\! 0.25\,$pc.  \citet{Fritz2011} concludes that little of the line of sight extinction towards the GC arises from within the central parsec, which we will assume to hold for sight-lines not passing through the major dust structures.  We also include a contribution from the much more extended old GC stellar component, using the radial profile from \citet{Fritz2014}, with a 3500~K spectrum normalized to $3 \!\times\! 10^7\,L_\odot$ within 100 arcsec, along with the uniform 2.73~K cosmic microwave background (CMB).

%--------------------------------------------------------------------%
\section{Geometry of Emission}
Assuming uniform emissivity, the flux arriving from a given direction can be calculated using ray tracing techniques.  For instance, we take an equation for a torus in Euclidean space, $f = (x^2 + y^2 + z^2 - r_1^2 - R_1^2)^2 + 4 R_1^2 (z^2 - r_1^2)$, insert the components for a ray ${\bold x}(t)$ starting from the electron position ${\bold r}_e$ and traversing direction ${\bold p}$, ${\bold x}(t) = {\bold r}_e + {\bold p}\, t$, and solve for the roots to find the length through ring 1, $\ell_1(\theta,\phi)$.  This involves solving a quartic equation, which can be done fairly quickly numerically.  The procedure either interior or exterior to spherical regions is similar.

\begin{figure}[t!]
%\hspace*{-0.0cm}
\includegraphics[width=1.01\columnwidth,clip=true]{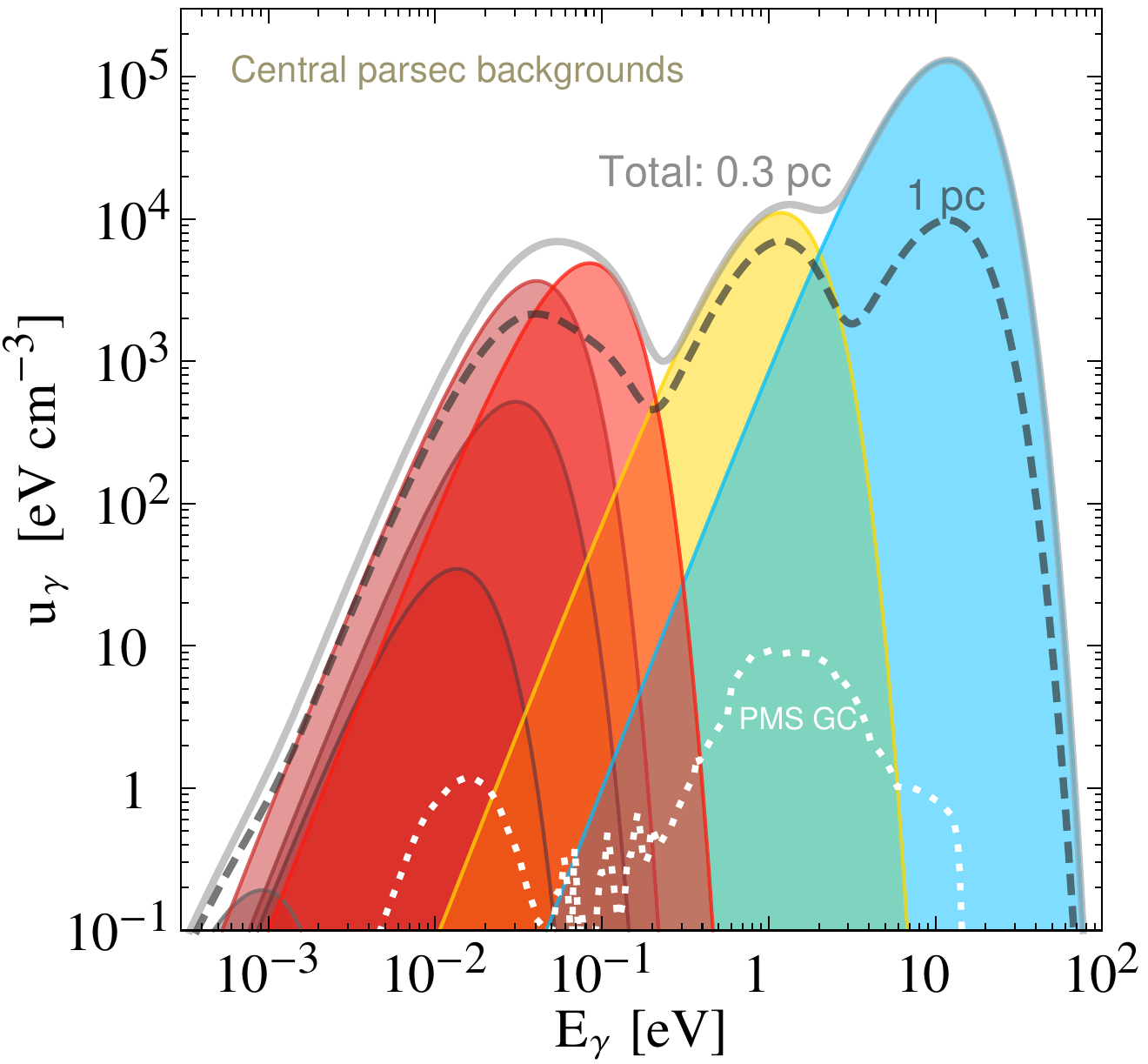}
\caption{Energy spectrum of background photons from our photon field.  Shown are the components of Table~\ref{tab:params} at a distance from Sgr~A$^*$ of 0.3~pc and their sum ({\it solid line}).  The {\it dashed line} shows the total background at 1~pc (in front and behind Sgr~A$^*$ are similar).  The GC background of \citet{Porter2006}, designed to be valid over larger scales, is also shown (PMS; {\it dotted}).\\
\label{SED}}
\end{figure}

To arrive at the energy density at a given position from each component, $u_i$, we do this many times en route to integrating over all angles
\begin{equation}
       u_i  =  \frac{L_{i}}{4\pi c\, V_i} \int d\Omega\, \ell_i(\theta,\phi)
    \,,
\label{ui}
\end{equation}
with $V_i$ the component volume.  Each spectral energy distribution is shown in Fig.~\ref{SED} along with the CMB (at $\sim\!10^{-3}\,$eV).

In Fig.~\ref{SED} we also compare to the oft-used modeled interstellar radiation field at the GC from \citet{Porter2006}.  Since this model is constructed from stellar contributions over larger scales, it is indicative of contributions within the central parsec from outside.  We see that our FIR energy density is larger by a factor of $\sim\! 10^3$ and so should remain dominant out to $\sim\!30\,$pc, corresponding to $\sim\!0.25^\circ$ (not accounting for any additional absorption).  Other more explicit contributions include the Arches and Quintuplet stellar clusters, which have luminosities comparable to the central cluster \citep{Figer2008}, but are relatively distant.  We thus assume these to be small in comparison to the local emission in what follows.

%%%%%%%%%%%%%%%%%%%%%%%%%%%%%%%%%%%
%%%%%%%%%%%%%%%%%%%%%%%%%%%%%%%%%%%

%--------------------------------------------------------------------%
\section{Gamma-ray Attenuation}
Our first application is to the attenuation of gamma rays due to $\gamma \gamma \rightarrow e^+ e^-$ interactions on intervening photon backgrounds.  The cross section depends on the relative angle with a gamma ray of energy $E_\gamma$ through $s \!=\! 2 E_\gamma \epsilon_\gamma (1-\cos{\theta})$ via $q \!=\! \sqrt{1-(2 m_e c^2)^2/s}$ as
\begin{equation}
       \sigma_{\gamma \gamma}(s)  \!=\!  \frac{3}{4} \sigma_T \frac{(m_e c^2)^2}{s} \left[ (3\!-\! q^4) \ln \!\frac{1\!+\! q}{1\!-\! q} \!-\! 2q (2\!-\! q^2)  \right]
    \!,
\label{sigmapair}
\end{equation}
with $\sigma_T$ the Thomson cross section.

\begin{figure}[b!]
\hspace*{-0.3cm}
%\vspace*{-0.175cm}
\includegraphics[width=1.05\columnwidth,clip=true]{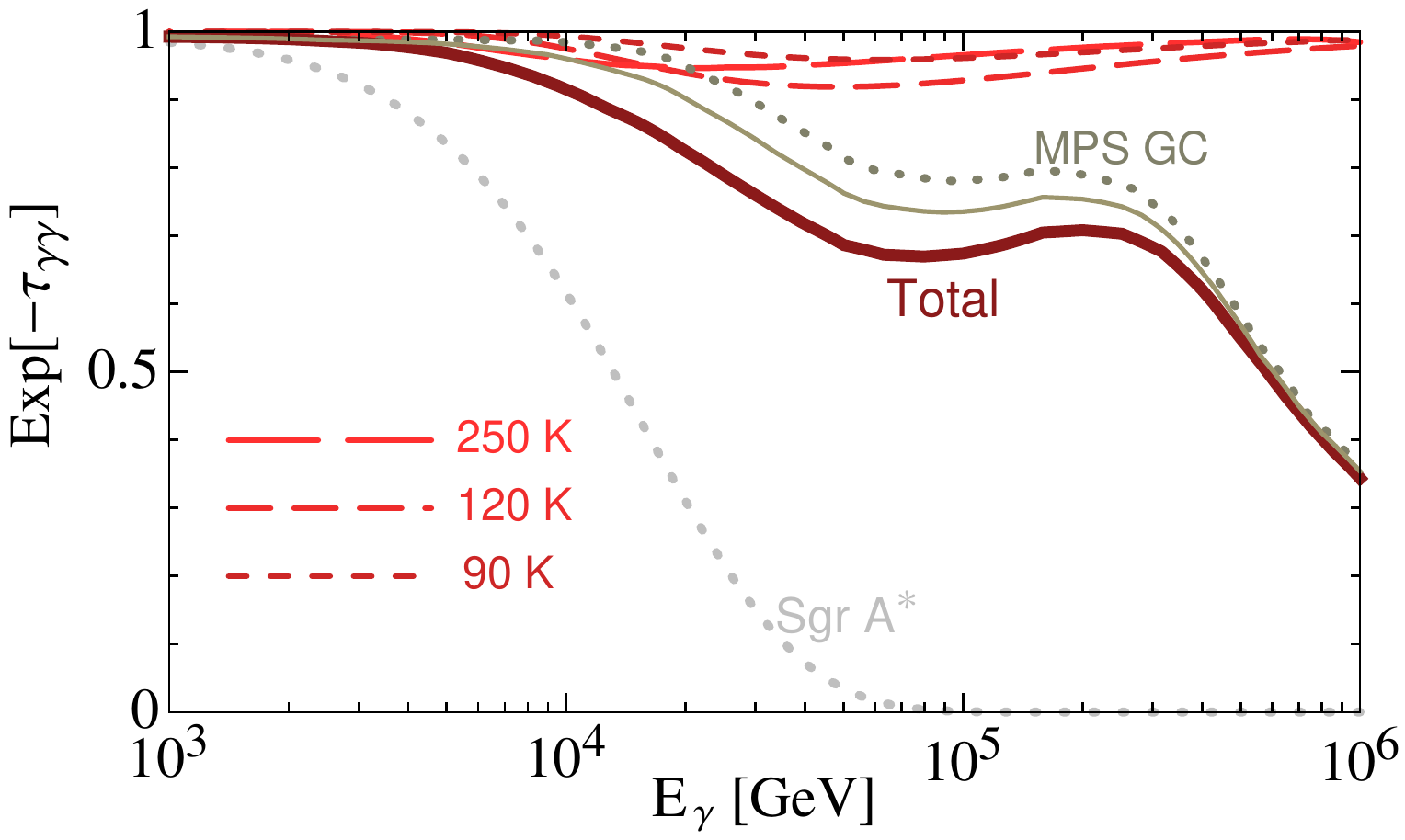}
\caption{Gamma-ray attenuation due to our background components from a location 1~pc behind the GC ({\it dashed}), the Galactic model of \citet{Moskalenko2006} (MPS; {\it dotted}), and their combination ({\it thick solid}).  Also, shown is the combined total from the GC position ({\it thin solid}), compared to attenuation within the inner accretion flow of Sgr~A$^*$ due to mm--IR emission ({\it dotted}).
\label{opa}}
\end{figure}

In Fig.~\ref{opa}, we show the result of integrating over two paths: one from the GC and a longer beam through the line of sight to PWN G359 (as denoted in Fig.~\ref{GCpc}) to 1~pc behind the GC.  Considering photon number density above the pair threshold, the 90~K, 120~K, and 250~K fields are the most important targets.  These are displayed for the latter case.  To obtain the total extinction, we add the GC attenuation curve from \citet{Moskalenko2006}, which is based on an interstellar radiation field model describing the galaxy on larger scales (plus the CMB), so double counting relative to our curves should be minimal.

We also display for comparison attenuation from within the inner accretion flow of Sgr~A$^*$.  We use spectra from \citet{Dexter2013} spanning from radio to IR (model 915h), with the simplifying assumption that this is spherical within a distance from the black hole of $3\, r_g$, with $r_g \!\simeq\! 6 \times 10^{11}$~cm, comparable to the IR emitting regions.  We see that this can be more important for any TeV gamma rays arising from within this limited volume around the black hole.

%--------------------------------------------------------------------%
\section{Electron Energy Loss Simply Stated}
We turn our attention to describing populations of electrons in the central parsec that can upscatter the above photon field into gamma rays.  We focus on energy spectra, not attempting to fully describe source morphology (though we remark on this later), evolving an injection spectrum with synchrotron and inverse Compton losses over a specified duration.  As far as X-rays from synchrotron are concerned, we will see that only the past few decades are relevant due to rapid cooling.

%
%%%%%%%%%%%%%%%%%%%%%%%%%%%%%
\begin{deluxetable}{rccc}[t!]
\tabletypesize{\scriptsize}
\tablecaption{\label{tab:params2}}
\tablewidth{\columnwidth}
\tablehead{\colhead{$T [K]$} & \colhead{$u_{\rm BB}$ [$10^{-9}\,$GeV~cm$^{-3}$]} & \colhead{$u_{i}/u_{\rm BB}$ (0.3~pc)} & \colhead{$u_{i}/u_{\rm BB}$ (1~pc)}  }
\startdata
\hline \vspace{-0.2cm}\\
%
% \hline
%
35000	& $7.1\!\times\!10^{15}$	& $2.5\!\times\!10^{-11}$	& $1.9\!\times\!10^{-12}$ \\
3500		& $7.1\!\times\!10^{11}$	& $2.1\!\times\!10^{-8}$	& $1.3\!\times\!10^{-8}$ \\
250		& $1.8\!\times\!10^{7}$	& $3.6\!\times\!10^{-4}$	& $8.3\!\times\!10^{-5}$ \\
120		& $9.8\!\times\!10^{5}$	& $5.1\!\times\!10^{-3}$	& $1.2\!\times\!10^{-3}$ \\
90		& $3.1\!\times\!10^{5}$	& $2.3\!\times\!10^{-3}$	& $4.0\!\times\!10^{-3}$ \\
40		& $1.2\!\times\!10^{4}$	& $3.9\!\times\!10^{-3}$	& $2.0\!\times\!10^{-3}$ \\
2.73		& $0.26$				& 1					& 1 \\
\hline\vspace{-0.3cm}
\enddata
\tablecomments{Energy density normalizations of the GC radiation components.\\}
\end{deluxetable}
%%%%%%%%%%%%%%%%%%%%%%%%%%%%%
%

%%%%%%%%%%%%%%%%%%%%%%%%%%%%%%%%%%%
%%%%%%%%%%%%%%%%%%%%%%%%%%%%%%%%%%%

Use of blackbody spectra allows for standard inverse Compton loss methods (dusty spectra will be examined elsewhere).  This can be done more or less exactly, although the resulting solution is rather cumbersome.  We rather examine first the form of the energy loss rate in the Thomson limit
\begin{equation}
       \left.\frac{dE_e}{dt}\right\vert_{\rm T}  =   - \frac{4}{3} \, \sigma_T \, c  \left(\frac{E_e}{m_e c^2}\right)^{\!2}  u_{\rm BB}
    \,,
\label{dEt}
\end{equation}
where $E_e$ is the electron energy, $m_e$ the electron mass, and $u_{\rm BB}$ the blackbody energy density for a given $T$, while in the extreme Klein-Nishina regime \citep{Blumenthal1970},
\begin{equation}
     \left.\frac{dE_e}{dt}\right\vert_{\rm KN} \! \!=\!   - \frac{\sigma_T}{16}  \frac{(m_e k_B T c)^2}{\hbar^3} \left( \ln 4 \kappa_e \!-\! 1.981 \right) \!,
\label{dEkn}
\end{equation}
where $\kappa_e \!=\! E_e k_B T/(m_e c^2)^2$.  To obtain $dE_e/dt\vert_{\rm IC}$ over the entire energy range, we find a convenient interpolation valid to $\sim\!1$\% below the KN limit,
\begin{equation}
  \! \!\!     \left.\frac{dE_e}{dt}\right\vert_{\rm H}  \!=\! -b_{\rm H} \kappa_e
       \left[\left(\frac{\kappa_e}{\kappa_1} \right)^{\!\!A \xi} \!\!+\! \left(\frac{\kappa_e}{\kappa_1} \right)^{\!\!B \xi}
        \!\!+\! \left(\frac{\kappa_2}{\kappa_1} \right)^{\!\!B \xi} \!\! \left(\frac{\kappa_e}{\kappa_2} \right)^{\!\!C \xi} \right]^{1/\xi} \!\!\!,
\label{hasanian}
\end{equation}
with $b_{\rm H} \!=\! 3.87 \!\times\! 10^{19}(k_B T)^2 \,$GeV$^{-1}$s$^{-1}$, $A \!=\! 1$, $B \!=\! -0.063$, $C \!=\! -0.855$, $\kappa_{1} \!=\! 0.065$, $\kappa_{2} \!=\! 4.16$, $\xi \!=\! -0.815$, and in which energy is given in terms of GeV.

Now we combine Eqs.~(\ref{dEkn}) and (\ref{hasanian}) as
\begin{equation}
    \!\!  \left.\frac{dE_e}{dt}\right\vert_{\rm IC} \! \!= \!  
        \left.\frac{dE_e}{dt}\right\vert_{\rm H} \Theta[10^{3.3} \!-\! \kappa_e]
        +\! \left.\frac{dE_e}{dt}\right\vert_{\rm KN} \Theta[\kappa_e \!-\! 10^{3.3}] 
        ,
\label{dEic}
\end{equation}
where $\Theta$ are step functions (cf., \citealt{Delahaye2010}).  This is to be evaluated for each distinct background component.

We also need consider the rate of energy loss due to synchrotron radiation,
\begin{equation}
       \left.\frac{dE_e}{dt}\right\vert_{\rm sync}  =   - \frac{4}{3} \, \sigma_T \, c  \left(\frac{E_e}{m_e c^2}\right)^2  u_B
    \,,
\label{dEtsync}
\end{equation}
for magnetic field energy density $u_B \!=\! B^2/8\pi$.  Adding this to the sum of the IC loss terms, we arrive at the total losses
\begin{equation}
       b_e(E_e)  = - \left.\frac{dE_e}{dt}\right\vert_{\rm sync} - \sum\limits_i \frac{u_i}{u_{\rm BB}} \left.\frac{dE_e}{dt}\right\vert_{{\rm IC,}\,i} 
    \,,
\label{dEtot}
\end{equation}
where each IC term is scaled by the ratio of the energy density of the photon background $u_i$ to the energy density of a pure blackbody $u_{\rm BB}$ for each $T_i$ (see Table~\ref{tab:params2}).

\begin{figure}[t!]\vspace*{-0.2cm}
\hspace*{-0.2cm}
\includegraphics[width=1.02\columnwidth,clip=true]{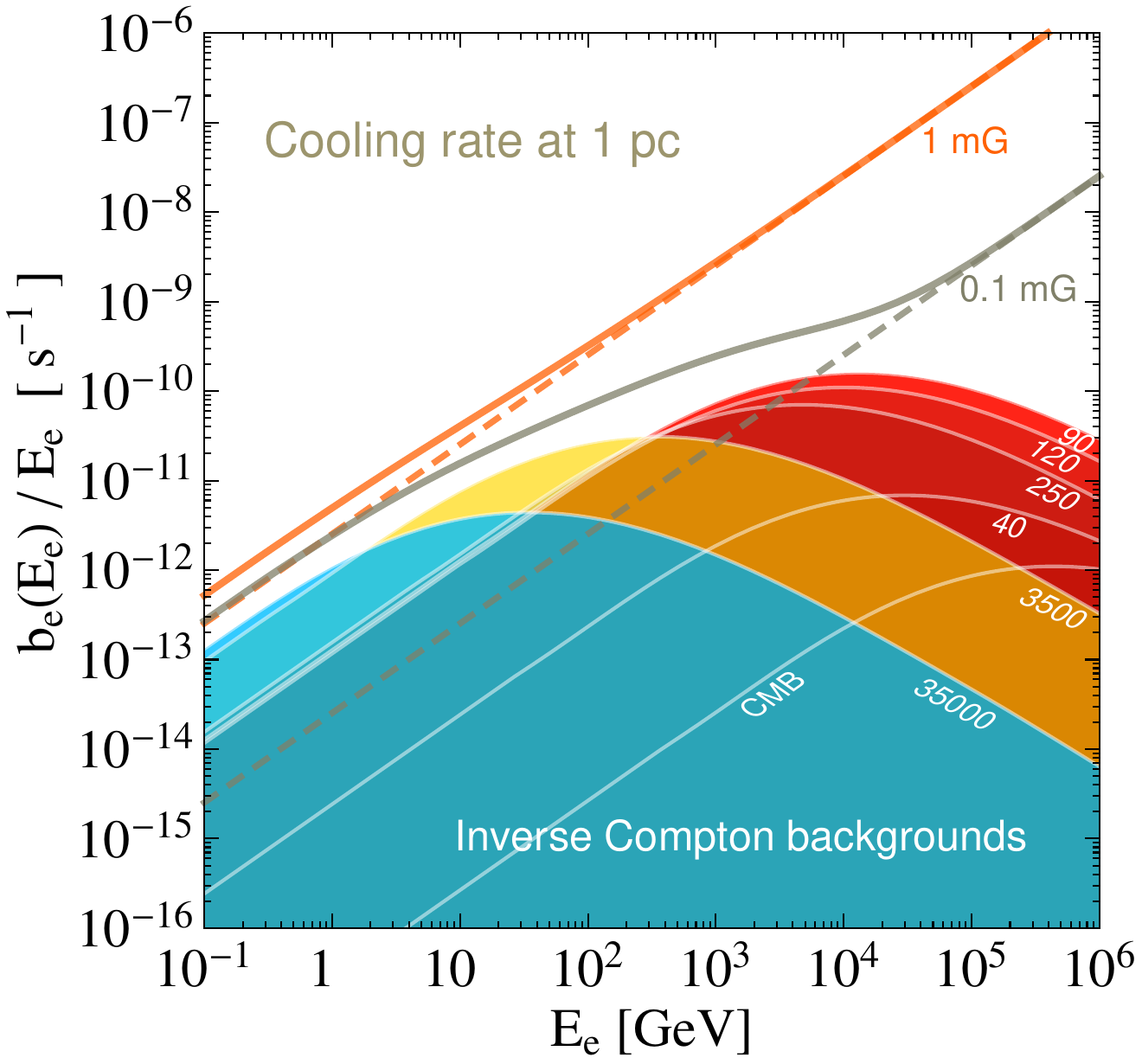}\vspace*{-0.15cm}
\caption{Rate of electron cooling due to synchrotron radiation for two field strengths ({\it dashed lines}), summed with inverse Compton losses on each background component of Fig.~\ref{SED} at a distance from Sgr~A$^*$ of 1~pc ({\it thin solid lines}; as labeled) to give the total loss rates ({\it thick solid lines}).\\
\label{bEfig}}
\end{figure}

In Fig.~\ref{bEfig}, we show the cooling rate, $b_e(E_e)/E_e$, for each component of the photon field at a distance of 1~pc from Sgr~A$^*$ and for two different choices of $B$ within the range discussed in \citet{Kistler2015} related to observations of the GC magnetar SGR J1745-29 \citep{Eatough2013}.  These demonstrate the change in relative importance of IC versus synchrotron as $B$ is varied as well as the KN suppression via the downturn in the IC curves, e.g., for $E_e \!\gtrsim\! 10$~TeV even the CMB is more relevant than the UV background.

%%%%%%%%%%%%%%%%%%%%%%%%%%%%%%%%%%%
%%%%%%%%%%%%%%%%%%%%%%%%%%%%%%%%%%%

%--------------------------------------------------------------------%
\section{Evolving the Electron Spectrum}
%\vspace*{0.1cm}
%
We are interested in the present population of electrons, which requires evolving the spectrum injected over all time.  To do so, we first determine the time it takes for an electron with initial energy $E_i$ to reach a final energy $E_f$ as
\begin{equation}
       t_l(E_i,E_f)  =   \int_{E_i}^{E_f} -\frac{dE}{b_e(E)} 
    \,.
\label{tloss}
\end{equation}
In practice, we take a very high energy, $E_h \!=\! 10^8\,$GeV, and evaluate $t_h(E_f) \!=\! t_l(E_h,E_f)$.  We then construct the inverse function $E_t[t_h(E_f)]$ numerically.  This is used as a convenient way to relate initial and final energies by taking the relative difference between them.  Now we integrate the source injection spectrum $dN_e/dE$ from a time $\tau$ up to today
\begin{equation}
        \frac{dN_e}{dE_0}  =   \int_{0}^{\tau} dt\, \left.\frac{dN_e}{dE\,dt}\right\vert_{E_t[t_h(E)-t]} \frac{b_e(E_t[t_h(E)-t])}{b_e(E)}
    \,.
\label{spec}
\end{equation}
This maps the source spectrum at each $t$ to the present time accounting for all relevant energy losses.

%--------------------------------------------------------------------%
\section{Synchrotron and Inverse-Compton Production}
%\vspace*{0.15cm}
%
We will consider a few illustrative problems of current interest, both in limits where synchrotron is dominant and where inverse Compton losses are clearly more important.  In evaluating the expected spectra of synchrotron and inverse Compton photons to compare with data, we assume that the electron population has a locally isotropic velocity distribution and that relativistic beaming effects are not relevant, though we do consider scattering off of anisotropic photon backgrounds.

Synchrotron can be elegantly calculated in the textbook manner using Bessel functions (e.g., \citealt{Rybicki1979}).  Since we are interested in an isotropic electron distribution, we instead follow the simpler approach in \citet{Aharonian2010}, with
\begin{equation}
      \frac{dN_\gamma}{dE_\gamma}  =   \frac{\sqrt{3}}{2 \pi} \frac{e^3 B}{m_e c^2 \hbar E_\gamma} G(x) \,e^{-x} \,,
\label{synch}
\end{equation}
where $x \!=\! 3 e \hbar B E_e^2/(2 m_e^3 c^5)$ and $G(x)$ is an interpolation close to the exact solution and faster to compute.  This is convolved with the present electron spectrum $dN_e/dE_0$.

Inverse Compton scattering becomes more involved, since we aim to examine the bulk angular dependence of central parsec photon backgrounds rather than assuming isotropy.  \citet{Khangulyan2014} provides a treatment convenient for this purpose (see also \citealt{Jones1968,Moskalenko2000,Zdziarski2013}).  For a mono-directional, blackbody photon distribution
\begin{equation}
      \frac{dN_{\rm ani}}{dE_\gamma}  \!=\!   \frac{3 \sigma_T \,m_e^2 c}{4 \pi^2 (\hbar c)^3} \frac{(k_B T)^2}{E_e^2} \! \left[ \frac{z^2}{2(1-z)} F_1(y) \!+\! F_2(y) \right]  \!,
\label{ani}
\end{equation}
with $z \!=\! E_\gamma/E_e$, $y \!=\! z (m_e c^2)^2/[2 (1-z) E_e k_B T (1 \!-\! \cos{\theta})]$.  Here the photons arrive at an angle $\theta$ to the electron, with the gamma ray departing in the electron direction.  Formulas for $F_1$ and $F_2$ are given in \citet{Khangulyan2014}, along with similar fitting equations $F_3$ and $F_4$ in case one is interested in using this technique to find the emission from an isotropic photon background, $dN_{\rm iso}/dE_\gamma$, e.g., the CMB.

%%%%%%%%%%%%%%%%%%%%%%%%%%%%%%%%%%%
\begin{figure*}[t!]%\vspace*{-0.25cm}
\hspace*{-0.2cm}
\includegraphics[width=0.6\columnwidth,clip=true]{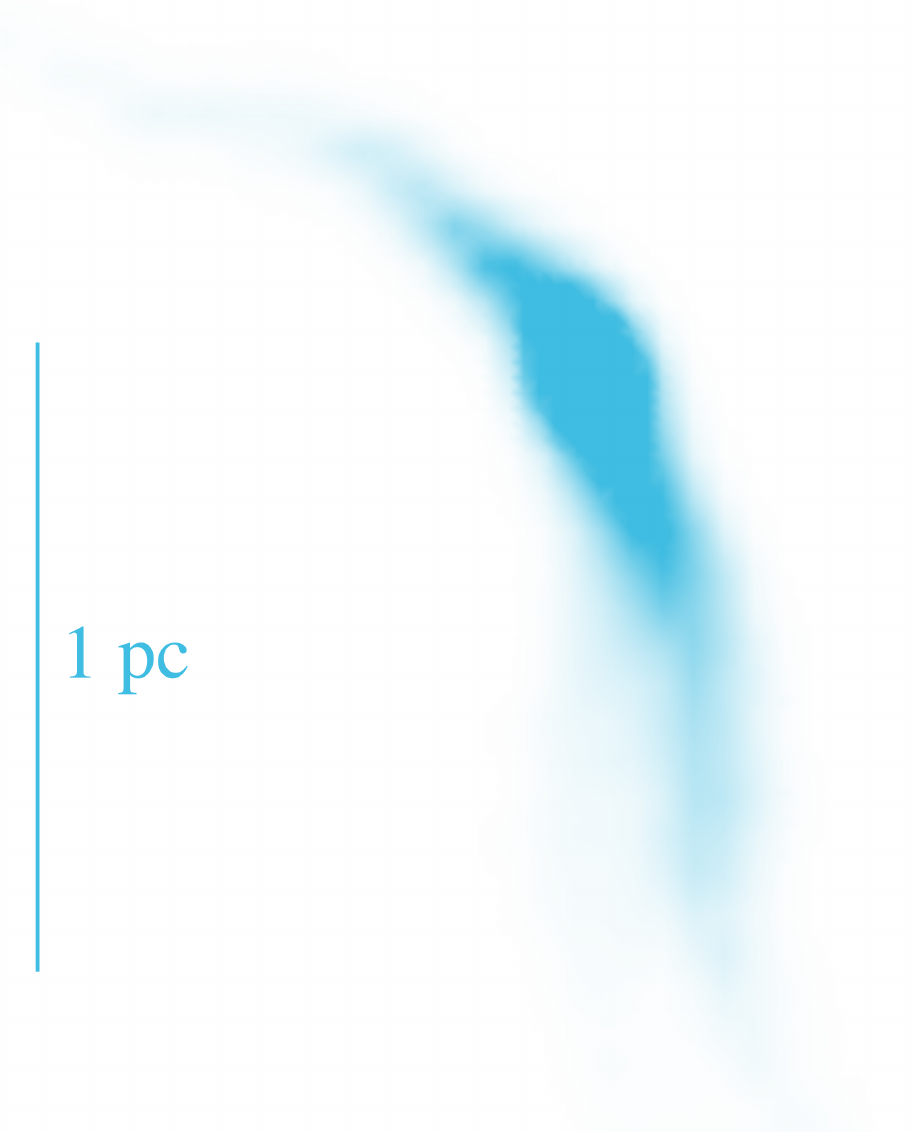}\hspace*{-0.5cm}
\includegraphics[width=1.6\columnwidth,clip=true]{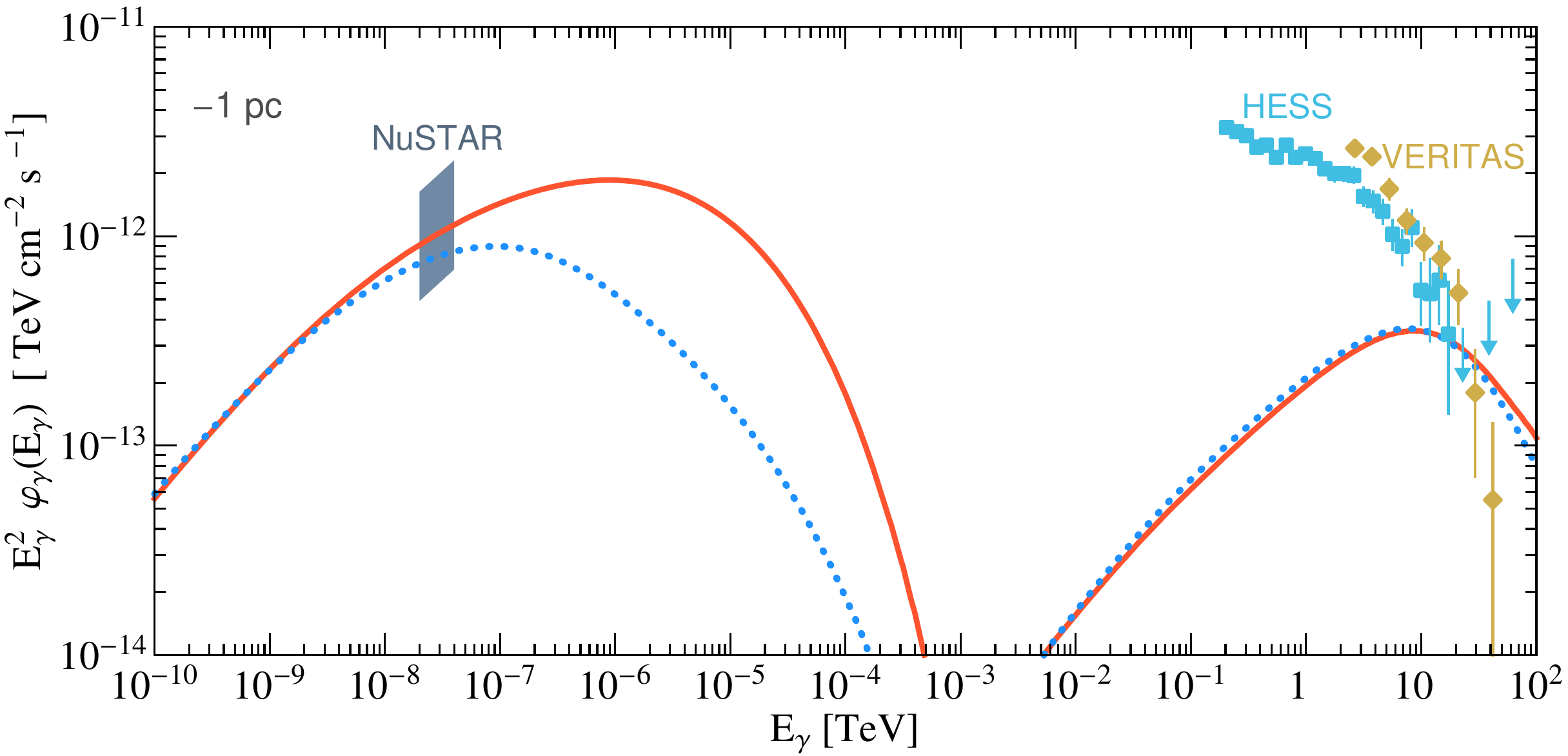}
\caption{{\it Left:} Projected distribution of electrons with 100~TeV initial energies continuously injected for 10~yr propagating in a 0.1~mG random magnetic field.
{\it Right:} Synchrotron and inverse Compton spectra from hard electron models in a 0.1~mG field with exponential ({\it solid lines}) and power-law ({\it dotted}) spectral breaks.  We show an approximate {\it NuSTAR} band and GC source TeV data from HESS \citep{Abramowski2016} and VERITAS \citep{Archer2016} for scale.\\
\label{stream}}
\end{figure*}
%%%%%%%%%%%%%%%%%%%%%%%%%%%%%%%%%%%

Using each angular-dependent photon field, we integrate from the source vantage point over angles with respect to the direction pointing at Earth to obtain the IC spectrum as
\begin{equation}
       \frac{dN_i}{dE_\gamma}  = \mathcal{E}_i  \int_{E_\gamma}^{E_{\rm max}} dE_e \frac{dN_e}{dE_0} \int d\Omega\, \frac{dN_{\rm ani}}{dE_\gamma} \ell_i(\theta,\phi)
    \,,
\label{ICspec}
\end{equation}
where $\mathcal{E}_i \!=\! L_i / (4\pi c\, u_{\rm BB} V_i)$.  One notable difference from assuming a central source is that the scattering on FIR emission from the rings is seen to vary much less in space.  We obtain fluxes $\varphi_i(E_\gamma)$ using a GC distance $d_{\rm GC} \!=\! 8.5\,$kpc.

%--------------------------------------------------------------------%
\section{Hard X-ray synchrotron and NuSTAR}
{\it NuSTAR} has recently discovered a diffuse hard X-ray flux reaching to $\gtrsim\!40\,$keV pervading the central parsecs \citep{Perez2015,Mori2015}.  While this could very well be due to some new class of sources endemic to the GC, synchrotron radiation is a well understood means of photon production, and while these X-ray energies are somewhat extreme, they are not terribly so and GC magnetic fields are unusually strong.

As Fig.~\ref{bEfig} shows, at sufficiently high energies synchrotron dominates over IC.  Examining the characteristic energy of synchrotron emission,
\begin{equation}
       E_\gamma \sim 20\, \left(\frac{E_e}{20\,{\rm TeV}}\right)^{\!2} \left(\frac{B}{{\rm mG}}\right) \, {\rm keV}
    \,,
\label{Echar}
\end{equation}
we see that hard X-rays can be the main product at these energies and field strengths.
Now, Fig.~\ref{bEfig} also shows that the cooling time (taking the inverse of the cooling rate) becomes quite short in this range, so one might expect X-ray emission to be limited to a small region around any such electron source.

However, as \citet{Giacinti2012} and \citet{Kistler2012} note, particles tend to propagate anisotropically at early times after injection from a fixed location, i.e., more quickly along the direction of the local magnetic field.  Following the arguments in \citet{Kistler2012}, if the cooling time is shorter than the characteristic timescale to reach isotropic diffusion, we would expect synchrotron emission to illuminate a path dependent on the local field structure since the particles only possess large energies for a limited duration.  If such a population is present in the GC, their bulk trajectories might be traceable by a hard X-ray telescope like {\it NuSTAR}.

To examine the plausibility of extended hard X-ray emission arising from electrons escaping a discrete source, we first consider the behavior of a population with initial energies of $\sim\!100\,$TeV.  Using the methods described in \citet{Kistler2012} and \citet{Yuksel2012}, we show in Fig.~\ref{stream} ({\it left}) an example of a possible realization of this scenario.  Here, we have injected 100~TeV electrons over a 0.1~pc radius volume in an isotropic random field configuration scaled to $B_{\rm rms} \!=\! 0.1\,$mG with a coherence length $l_c \!\approx\! 4\,$pc.  We inject continuously for $\sim\! 10\,$yr, over which time the energy can decrease to $\sim\! 50$~TeV.

While more elaborate simulations are possible, accounting for a spectrum of injected particles and energy dependence of propagation, this serves to illustrate the basic picture if high-energy electrons are not confined and free to propagate with only the local field guiding them, which may well be predominantly along the Galactic plane.  Alternatively, extended emission could arise from jet-like structures as seen reaching from some pulsars (e.g., IGR J11014--6103; \citealt{Pavan2014}).

%%%%%%%%%%%%%%%%%%%%%%%%%%%%%%%%%%%
\begin{figure*}[t!]
\hspace*{-0.2cm}
\includegraphics[width=2.12\columnwidth,clip=true]{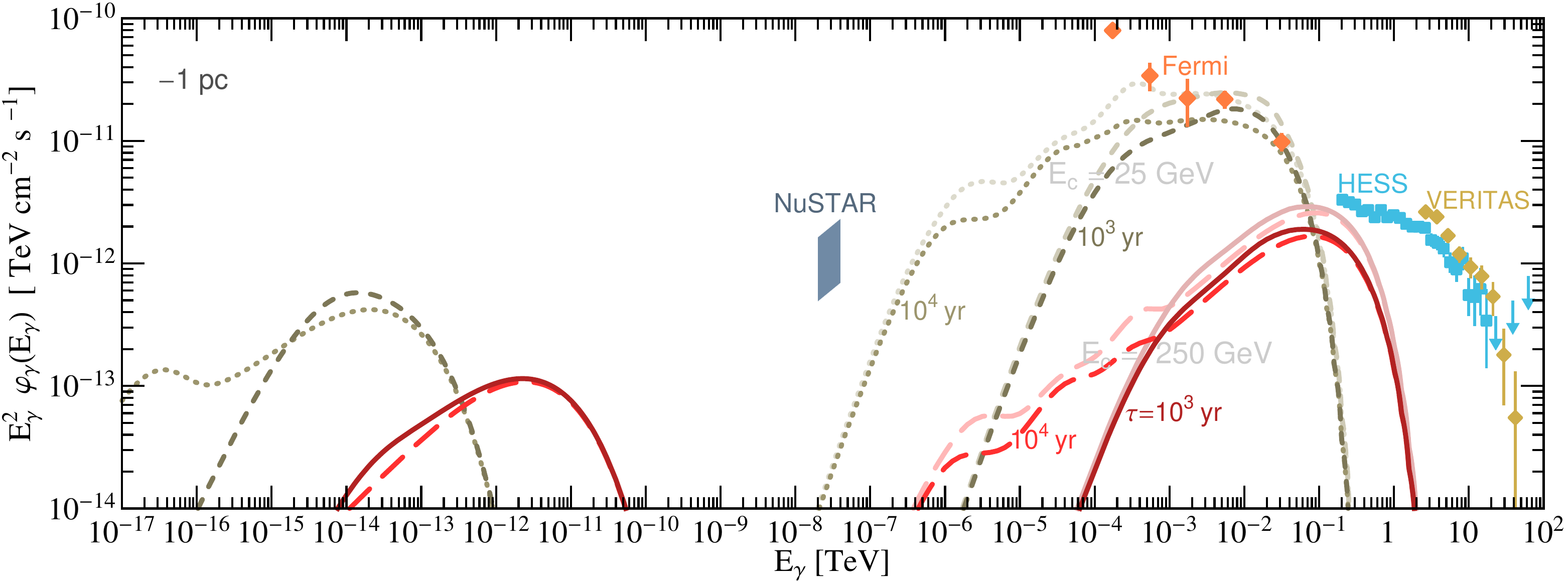}
\caption{Models in a 0.1~mG field to mimic a PWN relativistic Maxwellian, with $E_c \!=\! 25\,$GeV or $E_c \!=\! 250\,$GeV and durations $\tau \!=\! 10^3\,$yr and $10^4\,$yr yielding synchrotron (far left lines) and inverse Compton gamma rays ({\it darker lines:} isotropic IC at 1~pc; {\it lighter lines:} anisotropic IC 1~pc behind the GC).  We include here the {\it Fermi} GC source 3FGL J1745.6--2859c \citep{Acero2015}.\\
\label{hardx}}
\end{figure*}
%%%%%%%%%%%%%%%%%%%%%%%%%%%%%%%%%%%

If electrons are capable of retaining high energies over such distances near the GC, we can consider the emission from a single source.  We now calculate possible X-ray and gamma-ray fluxes using the methods described above.  \citet{Kistler2012} claims that while a bulk anisotropy may be present, the local velocity distribution can still be fairly isotropic.  In order to account for the hard spectrum of hard X-rays seen by {\it NuSTAR}, a hard electron spectrum may be needed.  We use a smoothly-broken power law with an exponential cutoff to describe the source spectrum
\begin{equation}
      \frac{dN_e}{dE}  =   f_e
       \left[\left(E/E_1\right)^{\alpha \eta} + \left(E/E_1\right)^{\beta \eta} \right]^{1/\eta} e^{-E/E_c}\,,
\label{fit}
\end{equation}
with $\alpha$ and $\beta$ the slopes, a break at $E_1$, cutoff energy $E_c$, and using $\eta \!=\! -10$ to give a sharp break.  We assume a constant injection spectrum and luminosity over a duration $\tau \!=\! 1000\,$yr.

We assume $\alpha \!=\! -1$ here.  Spectra as hard as this have been displayed recently in reconnection simulations (e.g., \citealt{Sironi2014}, \citealt{Guo2014}, \citealt{Werner2014}).  This is also representative of any harder spectra, since an equilibrium $\sim\! E_e^{-2}$ electron spectrum would generically result from continuous injection and cooling, leading to an X-ray spectrum of $\sim\! E_\gamma^{-1.5}$.  The spectral cutoff at low energies is not relevant here.  We consider cases where the high-energy break is due to an exponential cutoff alone at $E_c \!=\! 1000\,$TeV, with luminosity $\mathcal{L}_e \!=\! 2 \!\times\! 10^{35}\,$erg~s$^{-1}$, or a change in index to $\beta \!=\! -2.5$ at $E_1 \!=\! 200\,$TeV with $E_c \!=\! 2000\,$TeV and $\mathcal{L}_e \!=\! 10^{35}\,$erg~s$^{-1}$.

Fig.~\ref{stream} ({\it right}) shows the X-ray and gamma-ray fluxes for a uniform 0.1~mG field compared to an approximated band for {\it NuSTAR}.  For these hard spectra, there need not be bright radio emission.  We also see that the KN suppression leads to gamma rays principally from electrons with energies lower than that yielding synchrotron in the {\it NuSTAR} range.  In a weaker field, electrons would retain their energy longer.  They would more easily travel large distances and, for the same photon field, emit more gamma rays.  However, for a location beyond the central parsec, the photon background would be lower and the emission could remain below the HESS data.

As for where the electrons arise, the most likely culprits could be a pulsar associated with G359 or some heretofore unknown young pulsar with a velocity too low or local conditions otherwise unfavorable to yielding a prominent cometary nebula (see \citealt{Kistler2015}).  For this scenario we have assumed that the highest energy electrons are able to escape and freely propagate.  The physical conditions that might permit this would depend on the nature of the source, whether one or both of a linear accelerator setup by magnetic reconnection or Fermi shock acceleration is operating, and the magnetic field structure.  Excesses could be present close to the pulsar, where the field should be larger, or where a coherent PWN flow ends.  This basic setup can also be applied to a population of hard X-ray sources, such as fainter PWNe due a large number of active pulsars near the GC \citep{OLeary2015,OLeary2016}, which we defer to elsewhere.

%--------------------------------------------------------------------%
\section{Pulsar Exhaust and Fermi}
Often one simply imposes a sharp break in the electron injection spectrum at some low energy (as we just assumed above).  However, depending upon prevailing conditions, models that place the acceleration of particles at the termination shock in the pulsar wind can imply thermalization into a relativistic Maxwellian spectrum based on the bulk Lorentz factor of particles in the wind, with the shock energizing only some fraction of these into a power law component (e.g., \citealt{Amato2006,Sironi2013}).

If such an exhaust from the electron acceleration process is produced and goes somewhere, though, it should be emitting.  We examine two possible outcomes using unbroken $\alpha \!=\! 2$ spectra cutoff with $E_c \!=\! 25\,$GeV (corresponding to a bulk pre-shock wind Lorentz factor $\Gamma \!\sim\!5 \times 10^4$ and pair multiplicity $\mathcal{M} \!\sim\! 10^5$) and present luminosity $\mathcal{L}_0 \!=\! 10^{36}\,$erg~s$^{-1}$ or $E_c \!=\! 250\,$GeV ($\Gamma \!\sim\!5 \times 10^5$, $\mathcal{M} \!\sim\! 10^3$, and $\mathcal{L}_0 \!=\! 10^{35}\,$erg~s$^{-1}$).

Assuming a continuous luminosity, the equilibrium electron spectrum from this hard injected population will again tend toward $\sim\! E_e^{-2}$.  While a fixed $\mathcal{L}_e$ is reasonable for X-rays and TeV gamma rays due to the short cooling times of the emitting particles, at lower energies the accumulated spectrum may be enhanced by the pulsar spin down history.  We consider
\begin{equation}
      \mathcal{L}_e(t)  =   \mathcal{L}_0 \left[\frac{1+(\tau-t)/\tau_p}{1+\tau/\tau_p} \right]^{-\frac{n+1}{n-1}} ,
\label{spindown}
\end{equation}
where $\tau$ is the pulsar age, $\tau_p$ is a characteristic spin down time, and we use the canonical dipole $n \!=\! 3$ \citep{Gaensler2006}, although measured values for very young pulsars are often less than this (e.g., \citealt{Livingstone2011}) which would imply a different evolutionary history.  Our choices of $\tau_p \!=\! 10^3\,$yr and $\tau$, as well as $E_c$, are motivated to illustrate relations to gamma-ray data.

Fig.~\ref{hardx} shows the gamma-ray and synchrotron spectra assuming injection has occurred for $\tau \!=\! 10^3\,$yr or $10^4\,$yr.  The distance is fixed to 1~pc with a 0.1~mG field, although we note that the lower loss rates at these energies would likely result in most gamma rays being produced beyond a nominal PWN.  Increasing the injection duration has the effect of accumulating GeV electrons and pushing the sub-GeV gamma-ray flux upwards.  The lighter IC lines show an enhancement due to assuming anisotropic IC from 1~pc behind Sgr~A$^*$.

This flux is compared to the {\it Fermi} source coincident with the Galactic Center, 3FGL J1745.6--2859c, using data points from the 3FGL source catalog \citep{Acero2015}, which roughly split the previous 2FGL GC source (\citealt{Nolan2012}; cf., \citealt{Chernyakova2011,Abazajian2014}) into two distinct sources.  We also show the TeV data for scale, though one must keep in mind that there is a possible mismatch of spatial scales between the gamma-ray data sets.

Using a larger $\tau \!=\! 10^5\,$yr would decrease the energy at which particles accumulate to $\sim\,$100~MeV.  The IC flux could be increased into the {\it NuSTAR} range and continued to {\it INTEGRAL} energies \citep{Belanger2006}.  This is though, a rather long duration to expect a high luminosity from a lone pulsar.  To go farther back would also, considering typical densities in this region (e.g., \citealt{Ferriere2012,Linden2012,YusefZadeh2013}) necessitate accounting for ionization/Coulomb losses, which should overtake IC at some point and deplete the electron population at lower energies (see, e.g., Fig.~1 of \citealt{Hinton2007}).

The large number of massive stars in the GC implies an enhanced supernova rate and can lead to a typical interval between pulsar births of $\sim\!10^4$--$10^5\,$yr  \citep{Dexter2014,Eatough2015,OLeary2015,OLeary2016}.  Comparing the fluxes with varying injection periods shows the general behavior for a pulsar population.  Relic electrons from inactive pulsars will no longer contribute gamma rays since high-energy particles have lost energy.  For fixed luminosity, a higher spectral cutoff means fewer particles accumulating at lower energies with time (compare the two $E_c$ sets in Fig.~\ref{hardx}).  So while the high-energy range is more sensitive to a combination of cutoff and age, the flux of softer gamma rays depends less on age than the total number of electrons injected.

One might hope to use synchrotron to track these GeV particles and constrain the morphology.  Though the details will again depend upon the ambient magnetic field, as well as the spin down history of the pulsar, we can make a few rough estimates.  An isotropic diffusion coefficient of $D \!\sim\! 10^{26}\,$cm$^2\,$s$^{-1}$ implies a  distance scale of $(2D \tau)^{1/2} \!\sim\! 1\,$pc for $\tau \!=\! 10^3\,$yr.  Generally, both $D$ and $\mathcal{L}_{\rm sync}$ depend on $B$, with the morphology of the emission depending upon the magnetic field configuration and the photon field geometry.  We defer detailed examination of such variations to elsewhere.

%--------------------------------------------------------------------%
\section{Discussion and Conclusions}
\label{concl}
The properties of the Galactic Center can lead to unusual phenomena.  Consider if you will our two examples.  The former examines extremely high-energy electrons, yet results mostly in photons emitted with much lower energies than those from our later example that considers much lower energy electrons.  The nominal setup and parameter values appear physically plausible while leading to fluxes of hard X-rays and GeV gamma rays near the observed levels, illustrating the additional lengths yet required to understand this zone and its surroundings at high energies.

We have used a simplified model for the photon field of this complicated region as a starting point for better understanding the production of gamma rays from energetic electrons.  This also helps in determining the fate of the gamma rays, produced by whatever process imagined, that may be attenuated by the same photon backgrounds.  This explores a tractable middle ground between assuming isotropic backgrounds and following photons from the level of known stars to the heating and emission of dust.  The latter course is perhaps difficult, but not impossible (e.g., \citealt{Shcherbakov2014} considered the expected starlight background near the G2 object), and would aid in addressing the following additional implications.

%--------------------------------------------------------------------%
\subsection{Electrons and TeV gamma rays}
In the hard X-ray range there is a paucity of backgrounds as compared to lower energies, so that emission might be attributable to synchrotron radiation even in a complex region.  Evidence for electron acceleration to extremely high energies by pulsars includes the $\gtrsim\,$100~MeV flares from the Crab nebula ascribed to synchrotron from PeV electrons \citep{Abdo2011,Tavani2011,Arons2012,Cerutti2013} and signatures of multi-TeV electrons from pulsars in the solar neighborhood (see, e.g., \citealt{Yuksel2009,Kistler2009}).  The pulsar wind nebula G359.95--0.04 \citep{Wang2006,Muno2008} suggests such processes are active near the GC, which may also be quite relevant to TeV gamma-ray data (see \citealt{Kistler2015} for greater detail).

Our photon field model also allows for examination of another distinct scenario involving TeV gamma rays.  While the gamma-ray opacity along the sight lines examined in Fig.~\ref{opa} ended up not being overwhelming, this did not have to be the case.  A larger young stellar flux and/or a larger fraction of dust reprocessing, as may have been present in the past or in more active extragalactic central parsec regions, could easily lead to a more substantial suppression \citep{Kistler2015b}.

Any TeV gamma-ray source in this region produces an extended distribution of electrons and positrons due to $\gamma \gamma \rightarrow e^+ e^-$ on the photon field.  Comparing the {\it NuSTAR} and TeV energetics in Fig.~\ref{stream}, these roughly coincide.  Although our result suggests that such a process is not currently efficient in our GC, a sufficiently recent outburst of TeV gamma rays would have left an $e^\pm$ detritus yet emitting synchrotron.

%--------------------------------------------------------------------%
\subsection{More on Galactic Center Hard X-rays}
As a more general point regarding hard X-rays, while absorption is largely irrelevant \citep{Wilms2000}, the unusual gas streams in the central parsec could possess column depths sufficient to cause appreciable Thomson scattering.  If so, models of the gas density can be compared to X-ray maps to examine variations in intensity to determine the relative geometry of the X-ray emission and estimate the gas column.  This would help to clear up uncertainties over the nature of the CND, between high \citep{Christopher2005,Montero2009} and low \citep{RequenaTorres2012,Harada2015} inferred masses.

We also note that {\it NuSTAR} has detected non-thermal hard X-ray emission from the radio filament Sgr~A--E, suggesting that the spectrum could be accounted for via injection of electrons from an unknown PWN \citep{Zhang2014}.  Comparing to the better resolved radio images of Sgr~A--E \citep{Ho1985,YusefZadeh1987,Morris2014}, we see that the tail of PWN G359 extrapolates back to this general vicinity.  If related, this would imply a coherent structure of $\sim\! 10\,$pc, not unprecedented in the Milky Way (e.g., \citealt{Pavan2014}), just not obviously realizable near the GC.  This would require a rather low field strength for electrons to retain their energy until they reach the larger fields in the filament.

%--------------------------------------------------------------------%
\subsection{Moving Beyond the Center, Dark Matter, and Neutrinos}
We have focused on positions within the central parsec, since at larger distances the benefit of bright, compact infrared emission potentially producing an unusually large amount of IC losses in a small volume is lost.  The rather generic pulsar wind parameters used lead to a flux within range of {\it Fermi} data and allow further room for accommodation.  For instance, there may well be other pulsars in this area yielding GeV gamma rays, either pulsed or from a wind.   For a local supernova rate of $\sim\! 10^{-4}\,$yr$^{-1}$ these lead to overlapping contributions in the {\it Fermi} range, with burn off of electrons due to the steep rate of losses simplifying matters at higher energies.

Beyond the incentives to understand the novel astrophysics at the Galactic Center, there is also the quests for dark matter and neutrinos.  Of recent interest are claims of a significant excess of gamma rays at $\sim\! 1\!-\!10\,$GeV.  This may or may not be related to dark matter, but it does seem to originate at the Center, the IC scattering of electrons from annihilation or decay (cf., \citealt{Cholis2014}) is a direct application.  Improved understanding of the mechanism behind GC gamma rays will directly affect the expected flux of neutrinos (e.g., \citealt{Crocker2005,Kistler2006}) and whether the PeV neutrino seen from the vicinity of the GC by IceCube \citep{Aartsen2013} has a Galactic origin \citep{Kistler2015c}.

On larger scales, there should also be energetic electrons present from these and other processes.  While the concentrated UV emission most relevant in the central parsec will drop off rapidly, the old stellar component falls off less steeply so its contribution to IC will become relatively more important and may show up at lower gamma-ray energies (cf., \citealt{Abazajian2015}).  One can also consider the aforementioned Arches and Quintuplet stellar clusters, although these are rather young and lack the longer history of star formation present in the central parsec, possibly leading to fewer young pulsars.  They notably would also not contain a supermassive black hole.  Along with the central parsec, these could provide useful checks to discriminate between dark matter, pulsars, and diffuse cosmic-ray background contributions, details of which we will explore elsewhere.

%%---------------------------------------------------------------------%
\acknowledgments
We thank John Beacom, Jason Dexter, Ryan O'Leary, Troy Porter, and Hasan Yuksel for useful discussions and the hospitality of Brandt-Leland during the completion of this paper.
MDK acknowledges support provided by Department of Energy contract DE-AC02-76SF00515, and the KIPAC Kavli Fellowship made possible by The Kavli Foundation.

%---------------------------------------------------------------------%
%\textbf{References}
%\vspace*{-0.5cm}


\begin{thebibliography}{99}
%\vspace*{-0.75cm}

%\cite{Aartsen2013}
\bibitem[Aartsen et al.(2013)]{Aartsen2013}
  Aartsen, M.~G., et al.\ [IceCube Collaboration]
  2013a, \prl, 111, 021103
  %First Observation of PeV-Energy Neutrinos with IceCube

%\cite{Abazajian2014}
\bibitem[Abazajian et al.(2014)]{Abazajian2014}
  Abazajian, K.~N., Canac, N., Horiuchi, S., \& Kaplinghat, M.\
  2014, \prd, 90, 023526
  %Astrophysical and dark matter interpretations of extended gamma-ray emission from the Galactic Center

%\cite{Abazajian2015}
\bibitem[Abazajian et al.(2015)]{Abazajian2015}
  Abazajian, K.~N., Canac, N., Horiuchi, S., Kaplinghat, M., \& Kwa, A.\
  2015, JCAP, 7, 013
  %Discovery of a New Galactic Center Excess Consistent with Upscattered Starlight

%\cite{Abdo2011}
\bibitem[Abdo et al.(2011)]{Abdo2011}
  Abdo, A.~A., et al.\ [Fermi-LAT Collaboration]
  2011, Science, 331, 739
  %Gamma-Ray Flares from the Crab Nebula

%\cite{Abramowski2016}
\bibitem[Abramowski et al.(2016)]{Abramowski2016}
  Abramowski, A., et al.\ [HESS Collaboration]
  2016, Nature, 531, 476
  %Acceleration of petaelectronvolt protons in the Galactic Centre

%\cite{Acero2015}
\bibitem[Acero et al.(2015)]{Acero2015}
  Acero, F.., et al.\ [Fermi-LAT Collaboration]
  2015, \apjs, 218, 23
  %Fermi Large Area Telescope Third Source Catalog

%\cite{Aharonian2004}
\bibitem[Aharonian et al.(2004)]{Aharonian2004}
  Aharonian, F., et al.\ [HESS Collaboration]
  2004, \aap, 425, L13
  %Very high energy gamma rays from the direction of Sagittarius A*

%\cite{Aharonian2009}
\bibitem[Aharonian et al.(2009)]{Aharonian2009}
  Aharonian, F., et al.\ [HESS Collaboration]
  2009, \aap, 503, 817
  %Spectrum and variability of the Galactic center VHE gamma-ray source HESS J1745-290

%\cite{Aharonian2010}
\bibitem[Aharonian et al.(2010)]{Aharonian2010}
  Aharonian, F.~A., Kelner, S.~R., \& Prosekin, A.~Y.\
  2010, \prd, 82, 043002
  %Angular, spectral, and time distributions of highest energy protons and associated secondary gamma rays and neutrinos propagating through extragalactic magnetic and radiation fields

%\cite{Ahnen2016}
\bibitem[Ahnen et al.(2016)]{Ahnen2016} 
  Ahnen, M.~L., et al.\  [MAGIC Collaboration]
  %``Observations of Sagittarius A* during the pericenter passage of the G2 object with MAGIC,''
  2016, arXiv:1611.07095
  %%CITATION = ARXIV:1611.07095;%%

%\cite{Ajello2016}
\bibitem[Ajello et al.(2016)]{Ajello2016}
  Ajello, M., et al.\ [Fermi-LAT Collaboration]
  2016, \apj, 819, 44
  %Fermi-LAT Observations of High-Energy $\gamma$-Ray Emission Toward the Galactic Center

%\cite{Albert2006}
\bibitem[Albert et al.(2006)]{Albert2006}
  Albert, J., et al.\ [MAGIC Collaboration]
  2006, \apjl, 638, L101
  %Observation of Gamma Rays from the Galactic Center with the MAGIC Telescope

%\cite{Amato2006}
\bibitem[Amato \& Arons(2006)]{Amato2006}
  Amato, E., \& Arons, J.\
  2006, \apj, 653, 325
  %Heating and Nonthermal Particle Acceleration in Relativistic, Transverse Magnetosonic Shock Waves in Proton-Electron-Positron Plasmas

%\cite{Archer2014}
\bibitem[Archer et al.(2014)]{Archer2014}
  Archer, A., et al.\  [VERITAS Collaboration]
  2014, \apj, 790, 149
  %Very-high Energy Observations of the Galactic Center Region by VERITAS in 2010-2012

%\cite{Archer2016}
\bibitem[Archer et al.(2016)]{Archer2016}
  Archer, A., et al.\  [VERITAS Collaboration]
  2016, \apj, 821, 129
  %TeV Gamma-ray Observations of The Galactic Center Ridge By VERITAS

%\cite{Arons2012}
\bibitem[Arons(2012)]{Arons2012}
  Arons, J.\
  2012, \ssr, 173, 341
  %Pulsar Wind Nebulae as Cosmic Pevatrons: A Current Sheet's Tale

%\cite{Barriere2014}
\bibitem[Barriere et al.(2014)]{Barriere2014}
  Barri{\`e}re, N.~M., Tomsick, J.~A., Baganoff, F.~K., et al.\
  2014, \apj, 786, 46
  %NuSTAR Detection of High-energy X-Ray Emission and Rapid Variability from Sagittarius A* Flares

%\cite{Belanger2006}
\bibitem[Belanger et al.(2006)]{Belanger2006}
  B{\'e}langer, G., Goldwurm, A., Renaud, M., et al.\
  2006, \apj, 636, 275
  %A Persistent High-Energy Flux from the Heart of the Milky Way: INTEGRAL's View of the Galactic Center

%\cite{Blumenthal1970}
\bibitem[Blumenthal \& Gould(1970)]{Blumenthal1970}
  Blumenthal, G.~R., \& Gould, R.~J.\
  1970, \rmp, 42, 237
  %Bremsstrahlung, Synchrotron Radiation, and Compton Scattering of High-Energy Electrons Traversing Dilute Gases

%\cite{Calore2014}
\bibitem[Calore et al.(2015)]{Calore2014}
  Calore, F., Cholis, I., McCabe, C., \& Weniger, C.\
  2015, \prd, 91, 063003
  %A Tale of Tails: Dark Matter Interpretations of the Fermi GeV Excess in Light of Background Model Systematics

%\cite{Cerutti2013}
\bibitem[Cerutti et al.(2013)]{Cerutti2013}
  Cerutti, B., Werner, G.~R., Uzdensky, D.~A., \& Begelman, M.~C.\
  2013, \apj, 770, 147
  %Simulations of Particle Acceleration beyond the Classical Synchrotron Burnoff Limit in Magnetic Reconnection: An Explanation of the Crab Flares

%\cite{Chernyakova2011}
\bibitem[Chernyakova et al.(2011)]{Chernyakova2011}
  Chernyakova, M., Malyshev, D., Aharonian, F.~A., Crocker, R.~M., \& Jones, D.~I.\
  2011, \apj, 726, 60
  %The High-energy, Arcminute-scale Galactic Center Gamma-ray Source

%\cite{Cholis2014}
\bibitem[Cholis et al.(2015)]{Cholis2014}
  Cholis, I., Hooper, D., \& Linden, T.\
  2015, \prd, 91, 083507
  %A Critical Reevaluation of Radio Constraints on Annihilating Dark Matter

%\cite{Christopher2005}
\bibitem[Christopher et al.(2005)]{Christopher2005}
  Christopher, M.~H., Scoville, N.~Z., Stolovy, S.~R., \& Yun, M.~S.\
  2005, \apj, 622, 346
  %HCN and HCO+ Observations of the Galactic Circumnuclear Disk

%\cite{Crocker2005}
\bibitem[Crocker et al.(2005)]{Crocker2005}
  Crocker, R.~M., Melia, F., \& Volkas, R.~R.\
  2005, \apjl, 622, L37
  %Neutrinos from the Galactic Center in the Light of Its Gamma-Ray Detection at TeV Energy

%\cite{Davidson1992}
\bibitem[Davidson et al.(1992)]{Davidson1992}
  Davidson, J.~A., Werner, M.~W., Wu, X., et al.\
  1992, \apj, 387, 189
  %The luminosity of the Galactic center

%\cite{Daylan2014}
\bibitem[Daylan et al.(2014)]{Daylan2014}
  Daylan, T., Finkbeiner, D.~P., Hooper, D., Linden, T., Portillo, S.~K.~N., Rodd, N.~L., \& Slatyer, T.~R.
  2014, arXiv:1402.6703
  %The Characterization of the Gamma-Ray Signal from the Central Milky Way: A Compelling Case for Annihilating Dark Matter

%\cite{Delahaye2010}
\bibitem[Delahaye et al.(2010)]{Delahaye2010}
  Delahaye, T., Lavalle, J., Lineros, R., Donato, F., \& Fornengo, N.\
  2010, \aap, 524, A51
  %Galactic electrons and positrons at the Earth: new estimate of the primary and secondary fluxes

%\cite{Dexter2013}
\bibitem[Dexter \& Fragile(2013)]{Dexter2013}
  Dexter, J., \& Fragile, P.~C.\
  2013, \mnras, 432, 2252
  %Tilted black hole accretion disc models of Sagittarius A*: time-variable millimetre to near-infrared emission

%\cite{Dexter2014}
\bibitem[Dexter \& O'Leary(2014)]{Dexter2014}
  Dexter, J., \& O'Leary, R.~M.\
  2014, \apjl, 783, L7
  %The Peculiar Pulsar Population of the Central Parsec

%\cite{Draine2003}
\bibitem[Draine(2003)]{Draine2003}
  Draine, B.~T.\
  2003, \araa, 41, 241
  %Interstellar Dust Grains

%\cite{Eatough2013}
\bibitem[Eatough et al.(2013)]{Eatough2013}
  Eatough, R.~P., Falcke, H., Karuppusamy, R., et al.\
  2013, \nat, 501, 391
  %A strong magnetic field around the supermassive black hole at the centre of the Galaxy

%\cite{Eatough2015}
\bibitem[Eatough et al.(2015)]{Eatough2015}
  Eatough, R.~P., Lazio, T.~J.~W., Casanellas, J., et al.\
  2015, arXiv:1501.00281
  %Observing Radio Pulsars in the Galactic Centre with the Square Kilometre Array

%\cite{Etxaluze2011}
\bibitem[Etxaluze et al.(2011)]{Etxaluze2011}
  Etxaluze, M., Smith, H.~A., Tolls, V., Stark, A.~A., \& Gonz{\'a}lez-Alfonso, E.\
  2011, \aj, 142, 134
  %The Galactic Center in the Far-infrared

%\cite{FeldmeierKrause2015}
\bibitem[Feldmeier-Krause et al.(2015)]{FeldmeierKrause2015}
  Feldmeier-Krause, A.\, Neumayer, N., Schodel, R., et al.\
  2015, \aap, 584, A2
  %arXiv:1509.04707
  % KMOS view of the Galactic Centre I: Young Stars are centrally concentrated

%\cite{Ferriere2012}
\bibitem[Ferri{\`e}re(2012)]{Ferriere2012}
  Ferri{\`e}re, K.\
  2012, \aap, 540, A50
  %Interstellar gas within ~10 pc of Sagittarius A?

%\cite{Figer2008}
\bibitem[Figer(2008)]{Figer2008}
  Figer, D.~F.\
  2008, arXiv:0803.1619
  % Massive Star Formation in the Galactic Center

%\cite{Fritz2011}
\bibitem[Fritz et al.(2011)]{Fritz2011}
  Fritz, T.~K., Gillessen, S., Dodds-Eden, K., et al.\
  2011, \apj, 737, 73
  %Line Derived Infrared Extinction toward the Galactic Center

%\cite{Fritz2014}
\bibitem[Fritz et al.(2014)]{Fritz2014}
  Fritz, T.~K., Chatzopoulos, S., Gerhard, O., et al.\
  2014, arXiv:1406.7568
  %The Nuclear Cluster of the Milky Way: Total Mass and Luminosity (long version)

%\cite{Gaensler2006}
\bibitem[Gaensler \& Slane(2006)]{Gaensler2006}
  Gaensler, B.~M., \& Slane, P.~O.\
  2006, \araa, 44, 17
  %The Evolution and Structure of Pulsar Wind Nebulae

%\cite{Genzel2010}
\bibitem[Genzel et al.(2010)]{Genzel2010}
  Genzel, R., Eisenhauer, F., \& Gillessen, S.\
  2010, \rmp, 82, 3121
  %The Galactic Center massive black hole and nuclear star cluster

%\cite{Giacinti2012}
\bibitem[Giacinti et al.(2012)]{Giacinti2012}
  Giacinti, G., Kachelriess, M., \& Semikoz, D.~V.\
  2012, \prl, 108, 261101
  %Filamentary Diffusion of Cosmic Rays on Small Scales

%\cite{Goicoechea2013}
\bibitem[Goicoechea et al.(2013)]{Goicoechea2013}
  Goicoechea, J.~R., Etxaluze, M., Cernicharo, J., et al.\
  2013, \apjl, 769, L13
  %Herschel* Far-infrared Spectroscopy of the Galactic Center. Hot Molecular Gas: Shocks versus Radiation near Sgr A

%\cite{Guo2014}
\bibitem[Guo et al.(2014)]{Guo2014}
  Guo, F., Li, H., Daughton, W., \& Liu, Y.-H.\
  2014, \prl, 113, 155005
  %Formation of Hard Power-laws in the Energetic Particle Spectra Resulting from Relativistic Magnetic Reconnection

%\cite{Harada2015}
\bibitem[Harada et al.(2015)]{Harada2015}
  Harada, N., Riquelme, D., Viti, S., et al.\
  2015, \aap, 584, A102
  %arXiv:1510.02904
  %Chemical Features in the Circumnuclear Disk of the Galactic Center

%\cite{Hinton2007}
\bibitem[Hinton \& Aharonian(2007)]{Hinton2007}
  Hinton, J.~A., \& Aharonian, F.~A.\
  2007, \apj, 657, 302
  %Inverse Compton Scenarios for the TeV Gamma-Ray Emission of the Galactic Center

%\cite{Ho1985}
\bibitem[Ho et al.(1985)]{Ho1985}
  Ho, P.~T.~P., Jackson, J.~M., Barrett, A.~H., \& Armstrong, J.~T.\ 
  1985, \apj, 288, 575
  %Interactions between the continuum sources in the galactic center and their immediate molecular environment

%\cite{Jones1968}
\bibitem[Jones(1968)]{Jones1968}
  Jones, F.~C.\
  1968, Phys.\ Rev., 167, 1159
  %Calculated Spectrum of Inverse-Compton-Scattered Photons

%\cite{Khangulyan2014}
\bibitem[Khangulyan et al.(2014)]{Khangulyan2014}
  Khangulyan, D., Aharonian, F.~A., \& Kelner, S.~R.\
  2014, \apj, 783, 100
  %Simple Analytical Approximations for Treatment of Inverse Compton Scattering of Relativistic Electrons in the Blackbody Radiation Field

%\cite{Kistler2006}
\bibitem[Kistler \& Beacom(2006)]{Kistler2006}
  Kistler, M.~D., \& Beacom, J.~F.\
  2006, \prd, 74, 063007
  %``Guaranteed and prospective galactic TeV neutrino sources,''

%\cite{Kistler2009}
\bibitem[Kistler \& Yuksel(2009)]{Kistler2009}
  Kistler, M.~D., \& Yuksel, H.\
  2009, arXiv:0912.0264
  %New Constraints on the Highest-Energy Cosmic-Ray Electrons and Positrons

%\cite{Kistler2012}
\bibitem[Kistler et al.(2012)]{Kistler2012}
  Kistler, M.~D., Yuksel, H., \& Friedland, A.\
  2012, arXiv:1210.8180
    %``Galactic Streams of Cosmic-ray Electrons and Positrons,''

%\cite{Kistler2014}
\bibitem[Kistler et al.(2014)]{Kistler2014}
  Kistler, M.~D., Stanev, T., Yuksel, H.\
  2014, \prd, 90, 123006
  %Cosmic PeV neutrinos and the sources of ultrahigh energy protons

%\cite{Kistler2015}
\bibitem[Kistler(2015)]{Kistler2015}
  Kistler, M.~D.\
  2015, arXiv:1511.01159
  %``A Tale of Two Pulsars,''

%\cite{Kistler2015b}
\bibitem[Kistler(2015b)]{Kistler2015b}
  Kistler, M.~D.\
  2015, arXiv:1511.01530
  %``Problems and Prospects from a Flood of Extragalactic TeV Neutrinos in IceCube ,''

%\cite{Kistler2015c}
\bibitem[Kistler(2015c)]{Kistler2015c}
  Kistler, M.~D.\
  2015, arXiv:1511.05199
  %``On TeV Gamma Rays and the Search for Galactic Neutrinos,''

%\cite{Kistler2016}
\bibitem[Kistler \& Laha(2006)]{Kistler2016}
  Kistler, M.~D., \& Laha, R.\
  2016, arXiv:1605.08781
  %``Multi-PeV Signals from a New Astrophysical Neutrino Flux Beyond the Glashow Resonance,''

%\cite{Krabbe1995}
\bibitem[Krabbe et al.(1995)]{Krabbe1995}
  Krabbe, A., Genzel, R., Eckart, A., et al.\
  1995, \apjl, 447, L95
  %The Nuclear Cluster of the Milky Way: Star Formation and Velocity Dispersion in the Central 0.5 Parsec

%\cite{Lau2013}
\bibitem[Lau et al.(2013)]{Lau2013}
  Lau, R.~M., Herter, T.~L., Morris, M.~R., Becklin, E.~E., \& Adams, J.~D.\
  2013, \apj, 775, 37
  %SOFIA/FORCAST Imaging of the Circumnuclear Ring at the Galactic Center

%\cite{Linden2012}
\bibitem[Linden \& Profumo(2012)]{Linden2012}
  Linden, T., \& Profumo, S.\
  2012, \apj, 760, 23
  %Exploring the Nature of the Galactic Center gamma Ray Source with the Cherenkov Telescope Array

%\cite{Livingstone2011}
\bibitem[Livingstone et al.(2011)]{Livingstone2011}
  Livingstone, M.~A., Ng, C.-Y., Kaspi, V.~M., Gavriil, F.~P., \& Gotthelf, E.~V.\
  2011, \apj, 730, 66
  %Post-outburst Observations of the Magnetically Active Pulsar J1846-0258. A New Braking Index, Increased Timing Noise, and Radiative Recovery

%\cite{Mills2013}
\bibitem[Mills et al.(2013)]{Mills2013}
  Mills, E.~A.~C., G{\"u}sten, R., Requena-Torres, M.~A., \& Morris, M.~R.\
  2013, \apj, 779, 47
  %The Excitation of HCN and HCO+ in the Galactic Center Circumnuclear Disk

%\cite{Montero2009}
\bibitem[Montero-Casta{\~n}o et al.(2009)]{Montero2009}
  Montero-Casta{\~n}o, M., Herrnstein, R.~M., \& Ho, P.~T.~P.\
  2009, \apj, 695, 1477
  %Gas Infall Toward Sgr A* from the Clumpy Circumnuclear Disk

%\cite{Mori2015}
\bibitem[Mori et al.(2015)]{Mori2015}
  Mori, K., Hailey, C.~J., Krivonos, R., et al.\
  2015, arXiv:1510.04631
  %NuSTAR Hard X-ray Survey of the Galactic Center Region I: Hard X-ray Morphology and Spectroscopy of the Diffuse Emission

%\cite{Morris2014}
\bibitem[Morris et al.(2014)]{Morris2014}
  Morris, M.~R., Zhao, J.-H., \& Goss, W.~M.\
  2014, IAU Symposium, 303, 369
  %Nonthermal filamentary radio features within 20 pc of the Galactic center

%\cite{Moskalenko2000}
\bibitem[Moskalenko \& Strong(2000)]{Moskalenko2000}
  Moskalenko, I.~V., \& Strong, A.~W.\
  2000, \apj, 528, 357
  %Anisotropic Inverse Compton Scattering in the Galaxy

%\cite{Moskalenko2006}
\bibitem[Moskalenko et al.(2006)]{Moskalenko2006}
  Moskalenko, I.~V., Porter, T.~A., \& Strong, A.~W.\
  2006, \apjl, 640, L155
  %Attenuation of Very High Energy Gamma Rays by the Milky Way Interstellar Radiation Field

%\cite{Muno2008}
\bibitem[Muno et al.(2008)]{Muno2008}
  Muno, M.~P., Baganoff, F.~K., Brandt, W.~N., Morris, M.~R., \& Starck, J.-L.\
  2008, \apj, 673, 251
  %A Catalog of Diffuse X-Ray-emitting Features within 20 pc of Sagittarius A*: Twenty Pulsar Wind Nebulae?

%\cite{Nolan2012}
\bibitem[Nolan et al.(2012)]{Nolan2012}
  Nolan, P.~L., et al.\ [Fermi-LAT Collaboration]
  2012, \apjs, 199, 31
  %Fermi Large Area Telescope Second Source Catalog

%\cite{OLeary2015}
\bibitem[O'Leary et al.(2015)]{OLeary2015}
  O'Leary, R.~M., Kistler, M.~D., Kerr, M., \& Dexter, J.\
  2015, arXiv:1504.02477
  %Young Pulsars and the Galactic Center GeV Gamma-ray Excess

%\cite{OLeary2016}
\bibitem[O'Leary et al.(2016)]{OLeary2016}
  O'Leary, R.~M., Kistler, M.~D., Kerr, M., \& Dexter, J.\
  2016, arXiv:1601.05797
  %Young Pulsars and the Galactic Center GeV Gamma-ray Excess

%\cite{Pavan2014}
\bibitem[Pavan et al.(2014)]{Pavan2014}
  Pavan, L., Bordas, P., P{\"u}hlhofer, G., et al.\
  2014, \aap, 562, A122
  %The long helical jet of the Lighthouse nebula, IGR J11014-6103

%\cite{Perez2015}
\bibitem[Perez et al.(2015)]{Perez2015}
  Perez, K., Hailey, C.~J., Bauer, F.~E., et al.\
  2015, \nat, 520, 646
  %Extended hard-X-ray emission in the inner few parsecs of the Galaxy

%\cite{Porter2006}
\bibitem[Porter et al.(2006)]{Porter2006}
  Porter, T.~A., Moskalenko, I.~V., \& Strong, A.~W.\
  2006, \apjl, 648, L29
  %Inverse Compton Emission from Galactic Supernova Remnants: Effect of the Interstellar Radiation Field

%\cite{RequenaTorres2012}
\bibitem[Requena-Torres et al.(2012)]{RequenaTorres2012}
  Requena-Torres, M.~A., G{\"u}sten, R., Wei{\ss}, A., et al.\
  2012, \aap, 542, L21
  %GREAT confirms transient nature of the circum-nuclear disk

%\cite{Rybicki1979}
\bibitem[Rybicki \& Lightman(1979)]{Rybicki1979}
  Rybicki, G.~B., \& Lightman, A.~P.\
  1979, Radiation Processes in Astrophysics (New York: Wiley)
  %Radiative processes in astrophysics

%\cite{Shcherbakov2014}
\bibitem[Shcherbakov(2014)]{Shcherbakov2014}
  Shcherbakov, R.~V.\
  2014, \apj, 783, 31
  %The Properties and Fate of the Galactic Center G2 Cloud

%\cite{Sironi2013}
\bibitem[Sironi et al.(2013)]{Sironi2013}
  Sironi, L., Spitkovsky, A., \& Arons, J.\
  2013, \apj, 771, 54
  %The Maximum Energy of Accelerated Particles in Relativistic Collisionless Shocks

%\cite{Sironi2014}
\bibitem[Sironi \& Spitkovsky(2014)]{Sironi2014}
  Sironi, L., \& Spitkovsky, A.\
  2014, \apjl, 783, L21
  %Relativistic Reconnection: An Efficient Source of Non-thermal Particles

%\cite{Smith2014}
\bibitem[Smith \& Wardle(2014)]{Smith2014}
   Smith, I.~L., \& Wardle, M.\
   2014, \mnras, 437, 3159
  %An analysis of HCN observations of the Circumnuclear Disc at the Galactic Centre

%\cite{Stostad2015}
\bibitem[St{\o}stad et al.(2015)]{Stostad2015}
 St{\o}stad, M., Do, T., Murray, N., Lu, J. R., Yelda, S., \& Ghez, A.\
 2015, \apj, 808, 106
 %Mapping the Outer Edge of the Young Stellar Cluster in the Galactic Center

%\cite{Tavani2011}
\bibitem[Tavani et al.(2011)]{Tavani2011}
  Tavani, M., et al.\ [AGILE Collaboration]
  2011, Science, 331, 736
  %Discovery of Powerful Gamma-Ray Flares from the Crab Nebula

%\cite{Wang2006}
\bibitem[Wang et al.(2006)]{Wang2006}
  Wang, Q.~D., Lu, F.~J., \& Gotthelf, E.~V.\
  2006, \mnras, 367, 937
  %G359.95-0.04: an energetic pulsar candidate near Sgr A*

%\cite{Werner2014}
\bibitem[Werner et al.(2014)]{Werner2014}
  Werner, G.~R., Uzdensky, D.~A., Cerutti, B., Nalewajko, K., \& Begelman, M.~C.\
  2014, arXiv:1409.8262
  %The extent of power-law energy spectra in collisionless relativistic magnetic reconnection in pair plasmas

%\cite{Wilms2000}
\bibitem[Wilms et al.(2000)]{Wilms2000}
  Wilms, J., Allen, A., \& McCray, R.\
  2000, \apj, 542, 914
  %On the Absorption of X-Rays in the Interstellar Medium

%\cite{Yuksel2009}
\bibitem[Yuksel et al.(2009)]{Yuksel2009}
  Y{\"u}ksel, H.,  Kistler, M.~D., \& Stanev, T.\
  2009, \prl, 103, 051101
    %``TeV Gamma Rays from Geminga and the Origin of the GeV Positron Excess,''

%\cite{Yuksel2012}
\bibitem[Yuksel et al.(2012)]{Yuksel2012}
  Y{\"u}ksel, H., Stanev, T., Kistler, M.~D., \& Kronberg, P.~P.\
  2012, \apj, 758, 16
  %The Centaurus A Ultrahigh-energy Cosmic-Ray Excess and the Local Extragalactic Magnetic Field

%\cite{YusefZadeh1987}
\bibitem[Yusef-Zadeh \& Morris(1987)]{YusefZadeh1987}
  Yusef-Zadeh, F., \& Morris, M.\
  1987, \apj, 320, 545
  %Structural details of the Sagittarius A complex - Evidence for a large-scale poloidal magnetic field in the Galactic center region

%\cite{YusefZadeh2013}
\bibitem[Yusef-Zadeh et al.(2013)]{YusefZadeh2013}
  Yusef-Zadeh, F., Hewitt, J.~W., Wardle, M., et al.\
  2013, \apj, 762, 33
  %Interacting Cosmic Rays with Molecular Clouds: A Bremsstrahlung Origin of Diffuse High-energy Emission from the Inner 2x1deg of the Galactic Center

%\cite{Zdziarski2013}
\bibitem[Zdziarski \& Pjanka(2013)]{Zdziarski2013}
  Zdziarski, A.~A., \& Pjanka, P.\
  2013, \mnras, 436, 2950
  %Compton scattering of blackbody photons by relativistic electrons

%\cite{Zhang2014}
\bibitem[Zhang et al.(2014)]{Zhang2014}
  Zhang, S., Hailey, C.~J., Baganoff, F.~K., et al.\
  2014, \apj, 784, 6
  %High-energy X-Ray Detection of G359.89-0.08 (Sgr A-E): Magnetic Flux Tube Emission Powered by Cosmic Rays?

%\cite{Zhao2009}
\bibitem[Zhao et al.(2009)]{Zhao2009}
  Zhao, J.-H., Morris, M.~R., Goss, W.~M., \& An, T.\
  2009, \apj, 699, 186
  %Dynamics of Ionized Gas at the Galactic Center: Very Large Array Observations of the Three-dimensional Velocity Field and Location of the Ionized Streams in Sagittarius A West

%\cite{Zhao2010}
\bibitem[Zhao et al.(2010)]{Zhao2010}
  Zhao, J.-H., Blundell, R., Moran, J.~M., et al.\
  2010, \apj, 723, 1097
  %The High-density Ionized Gas in the Central Parsec of the Galaxy


\end{thebibliography}
\end{document}